\documentclass[12pt]{article}
 
\usepackage{gensymb}
\usepackage{url}
\usepackage{amssymb,amsmath, amsfonts}
\usepackage{cases}
\usepackage{latexsym}
\usepackage{graphicx}
\usepackage{relsize}
\usepackage{natbib}
\usepackage{rotating}
\usepackage[includeheadfoot,left=1.in,right=1.in,top=1.in,bottom=1.in,bindingoffset=0.in,nohead]{geometry}
\usepackage[onehalfspacing]{setspace}
\usepackage{booktabs}
\usepackage{appendix} 
\usepackage[pdfborder={0 0 0}]{hyperref} 
\usepackage{float}
\usepackage{xcolor}  
\usepackage{caption}
\usepackage{subcaption}
\usepackage{minibox}
\usepackage{enumitem}

\begin{document}

	\title{Real-Time Real Economic Activity: \\Entering and Exiting the Pandemic Recession of 2020  \\  \bigskip}

%	\author{Francis X. Diebold\\University of Pennsylvania \\$~$\\$~$\\First Version:  June 26, 2020\\This Version, \today}
		
			\author{Francis X. Diebold\\University of Pennsylvania \\$~$\\ \today}
			
		\date{}
		
	\maketitle

\bigskip

	\begin{spacing}{1}

		\noindent  \textbf{Abstract}: Entering and exiting the Pandemic Recession, I study the high-frequency real-activity signals provided   by  a leading nowcast, the ADS Index of Business Conditions produced and released in real time  by the Federal Reserve Bank of Philadelphia.  I track the evolution of real-time vintage beliefs  and compare them to a  later-vintage  chronology.   Real-time ADS plunges and then swings as its underlying economic indicators swing, but the ADS paths quickly converge to indicate a return to brisk positive growth by mid-May. I show, moreover, that  the daily real activity path was highly correlated with the daily COVID-19 cases.  Finally, I provide a comparative assessment of  the real-time ADS signals provided when exiting the Great Recession.

		\thispagestyle{empty}
				
		\bigskip

		\bigskip
	
		\bigskip
		
		\noindent {\bf Acknowledgments}:  This paper is a revised and extended version of \cite{Diebold2020}.  For helpful discussion I thank (without implicating) Boragan Aruoba, Scott Brave, Daniel Lewis, Tony Liu, Andrew Patton, Glenn Rudebusch, Frank Schorfheide, Chiara Scotti, Minchul Shin, Keith Sill, Jim Stock, Mark Watson, Simon van Norden, and especially Tom Stark.  For outstanding research assistance and related discussion I thank Philippe Goulet Coulombe, Boyan Zhang, and especially Tony Liu.  The usual disclaimer applies.

		\bigskip


		\bigskip


		\bigskip

		\noindent {\bf Contact}:  fdiebold@sas.upenn.edu

	\end{spacing}

	\clearpage
%	
%	\thispagestyle{empty}
%	\setcounter{tocdepth}{3}
%	\tableofcontents
%	
%	
%	
%	\clearpage
%	
	\setcounter{page}{1}
	\thispagestyle{empty}

\section{Introduction}

Accurate assessment of of current real economic activity (``business conditions") is key for successful decision making in business, finance, and policy.  It is difficult, however,  to track business conditions in real time, both because no single observed economic indicator \textit{is} ``business conditions", and  because different indicators are available at different observational frequencies, and with different release delays. Nevertheless there exists the tantalizing possibility of accurate real-time business conditions assessment (``nowcasting"), and recent decades have witnessed great interest in nowcasting methods and  applications (e.g., \cite{Banburaetal2011}).

The workhorse nowcasting approaches involve dynamic factor models, which relate a set of observed real activity indicators to a single underlying latent real activity factor.  Both ``small data" approaches (e.g., based on 5 indicators) and ``Big Data" approaches (e.g., based on 500 indicators)  are available.  Small data approaches were developed first, and they typically involve maximum likelihood estimation (e.g., \cite{SW1989}). Subsequent Big Data approaches, in contrast, typically involve two-step estimation based on a first-step extraction of principal components (e.g., \cite{SW2002}, \cite{FRED2016}). 

Both introspection and experience reveal that  Big Data nowcasting approaches are not necessarily better.  First, they are more tedious to manage, and less transparent. Second, they may not deliver much improvement in factor extraction accuracy, which increases  and stabilizes quickly as the number of indicators increases  \citep{DGR2012}. Third, casual inclusion of many indicators can be problematic because a poorly-balanced set of indicators can create distortions in the extracted factor \citep{BoivinNg2006}, whereas small data approaches promote and facilitate hard thinking about a well-balanced set of indicators (\cite{BaiandNg2008}).

Against this background, in this paper I assess the performance of a leading small-data nowcast, the Aruoba-Diebold-Scotti (ADS) Index of Business Conditions \citep{ADS2009}.  ADS is designed to track real business conditions at high frequency, and it has been maintained and released in real time by the Federal Reserve Bank of Philadelphia continuously since 2008.\footnote{The production version  used by FRB Philadelphia differs in some ways (e.g., included indicators and treatment of trend) from the prototypes  provided by \cite{ADS2009} and \cite{AD2010}, which themselves differ slightly.  All discussion in this paper refers to the FRB Philadelphia version.  All materials, including the full set of vintage nowcasts, are available at \url{https://www.philadelphiafed.org/research-and-data/real-time-center/business-conditions-index}.}    Its modeling style and underlying economic indicators  build on classic early work in the tradition of  \cite{BM1946}, \cite{SargentandSims1977}, and \cite{SW1989}.  The underlying indicators span high- and low-frequency information on real economic flows:  weekly initial jobless claims; monthly payroll employment growth, industrial production growth, personal income less transfer payments growth, manufacturing and trade sales growth; and quarterly real GDP growth.

Crucially,  I  assess ADS  using only information actually available in real time.  This is  required for truly credible real-time evaluation, and it  can only be achieved by using nowcasts produced and permanently recorded in real time, which is very different from simply removing final-revised data and inserting vintage data into an otherwise ex post analysis.  Unfortunately,  such  evaluations are rare, because there simply are not many instances of long series of nowcasts produced and recorded in real time.  ADS, however, has been produced and recorded in real time  roughly twice weekly since late 2008, so  I  can provide real-time performance assessments both exiting the Great Recession and entering/exiting the Pandemic Recession.  

Ultimately the paper takes a two-pronged approach.  The first is the above-sketched  attempt at real-time ADS assessment, asking whether ADS sends  reliable signals. The second conditions on reliability of the signals, and uses them to assess what actually happened in the Pandemic Recession of 2020 (and, for comparison, in the Great Recession of 2007-2009).  The two prongs are ultimately inseparable and woven together in various ways throughout the paper.

 I   proceed as follows. In section \ref{constr},  I  provide background on aspects of ADS construction, updating, ex post characteristics, and  performance evaluation.  In section \ref{PR},  I   examine  ADS entering/exiting the Pandemic Recession, and  I  relate the real-time ADS path to the real-time COVID-19 path. In section \ref{GR}, I provide a comparative examination of ADS exiting the Great Recession.   I  conclude in section \ref{concl}.

\section{Nowcast  Construction,  Characteristics, and  Assessment} \label{constr}

Here  I  provide background on the ADS index construction (section \ref{aa1}),  ex post historical characteristics  (section \ref{aa2}), and general  issues of relevance to assessing ex ante nowcasting performance  (section \ref{aa3}).

\subsection{Construction and Updating} \label{aa1}

ADS is a dynamic factor model with multiple mixed-frequency real activity indicators driven by a single latent real activity factor. The ADS index is an estimate of that latent real activity factor. Importantly, the model is specified such that \textit{the real activity factor tracks the de-meaned growth rate of real activity}.  Progressively more negative or positive values indicate progressively worse- or better-than-average real growth, respectively. Because ADS tracks real activity growth, not level, a positive value does not necessarily mean ``good times"; rather, it means ``good growth", which may be from a level well below trend, as for example in the early stages of a recovery.

ADS is specified at daily frequency, allowing as necessary for missing data for the less-frequently observed variables.\footnote{The model must be specified at daily frequency, despite the fact that the  highest-frequency indicator is is weekly initial jobless claims, to account for the varying number of days/weeks per month, which also produces time-varying system parameter matrices.}  Importantly, despite complications from  missing data, time-varying system matrices, aggregation across frequencies, etc.,  the  Kalman filter and associated Gaussian pseudo likelihood evaluation via prediction-error decomposition remain valid, subject to some well-known modifications.\footnote{See, for example, \cite{DurbinandKoopman2001} on missing data, and  \cite{Harvey1991} on aggregation of flow variables.}  Model estimation is therefore straightforward, after which the Kalman smoother produces an optimal extraction of the underlying real activity factor. That is, the Kalman smoother produces the index: The  extracted sequence at any time $t^*$ \textit{is} the vintage-$t^*$ ADS sequence, $\{ ADS_1, ADS_2,  ...ADS_{t^*} \}$.  

The first ADS vintage was released  12/5/2008, covering 3/1/1960 through 11/30/2008.  Since then, ADS has been continuously updated whenever new data are released.  The Kalman smoother is re-run, generally within two hours of the release,  and the newly-extracted index from  3/1/1960 to ``the present" is re-written to the web.  ADS has been updated  approximately eight times per month on average since inception.

\subsection{Ex Post  Characteristics}  \label{aa2}

\begin{figure}[tbp]
	\caption{ADS Index:  Ex Post Path  03/01/1960 - 12/31/2013 (Vintage 6/26/2020)}
	\begin{center}
		\includegraphics[trim={40mm 20mm 0 25mm},clip,scale=.29]{{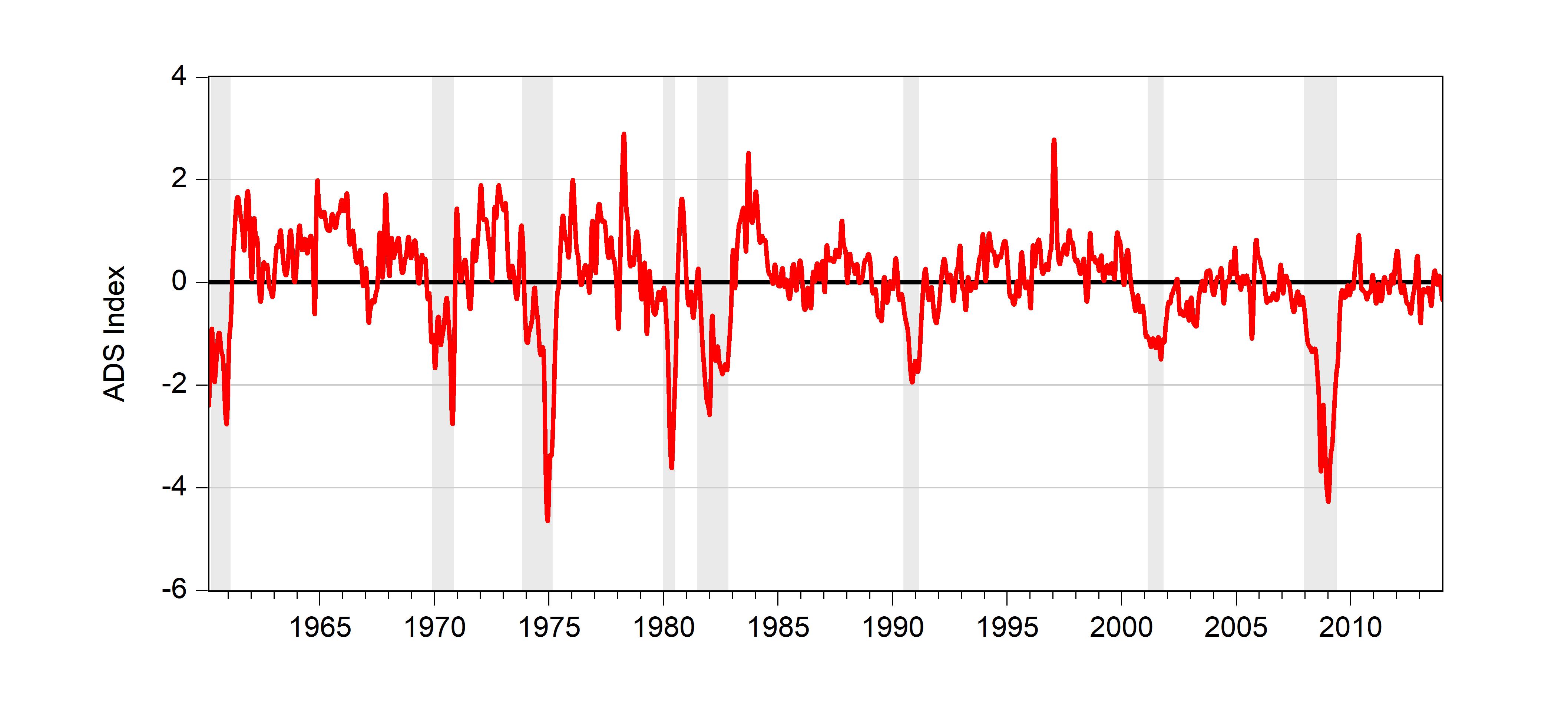}}
	\end{center}
	\label{ExPost}
	Notes:   The shaded regions are NBER-designated recessions.
\end{figure}

In Figure \ref{ExPost}  I  show the ADS index from 03/01/1960 through 12/31/2013, as assessed in the 6/26/2020 vintage.  The sample range is well before the vintage pull date, so the chronology displayed is  (intentionally) ex post.  I  do this because it is instructive to examine the ex post chronology before passing to real time assessment, which can only be done after ADS went live in late 2008.\footnote{The sample period intentionally excludes the Pandemic Recession, which  I  will subsequently examine in detail.}

Several features are noteworthy.  For example,  the ADS chronology coheres strongly with the NBER chronology, plunging during NBER recessions.   In addition,  several often-discussed features of the business-cycle are evident in ADS, such as   the pronounced moderation in volatility during the Greenspan era.

The ADS value added relative to the NBER chronology stems from the facts that (1) it is a cardinal measure, allowing one to assess not only recession durations, but also depths and patterns (see Table \ref{NBERreplica}), and (2) its updates arrive  in timely fashion, whereas the starting and ending dates NBER recessions  are typically not announced until well after the fact (again see Table \ref{NBERreplica}).  Of course, if ADS is  to be a useful guide for business and policy decisions,  its frequently-arriving updates must provide reliable signals in real time, not just ex post as in Figure \ref{ExPost}.   I  now turn to that issue.  

\begin{table}[tb]
	\caption{NBER Recessions}
	\label{NBERreplica}
	\begin{center}
		\begin{tabular}{lllcc}
			\toprule
			\multicolumn{2}{c}{Recession  Dates } & \multicolumn{3}{c}{Recession Characteristics} \\
			\midrule
			\midrule
			Peak Month  & Trough Month & \multicolumn{1}{l}{Duration} & \multicolumn{1}{l}{Depth}  & \multicolumn{1}{l}{Severity}  \\
			\midrule
			April 1960 & February 1961 & 10           &  2.7 & 27.0 \\
			December 1969 & November 1970 & 11          &  2.8  & 30.8 \\
			November 1973 & March 1975 & 16            &  4.7 & 75.2\\
			January 1980 (6/3/1980) & July 1980 (7/8/1981) & 6         &   3.6 &  21.6  \\
			July 1981 (1/6/1982) & November 1982 (7/8/1983) & 16     &   2.9  &  46.4 \\
			July 1990 (4/25/1991) & March 1991 (12/22/1992) & 8      &  1.7   &  13.6  \\
			March 2001 (11/26/2001) & November 2001 (7/17/2003) & 8       &   1.5 &  12.0  \\
			December 2007 (12/1/2008) & June 2009 (9/20/2010) & 18   &  4.3  & 77.4 \\
			February 2020 (6/6/2020)  &  April 2020 (7/19/2021) & 2   & 26.6   & 53.2   \\
			\bottomrule
		\end{tabular}
	\end{center}
	\begin{spacing}{1.0} \footnotesize \noindent  Notes:  Recession dates and durations in months are from the NBER chronology; see \url{https://www.nber.org/cycles.html}.  Announcement dates appear in parentheses.   Recession depth is the minimum absolute daily ADS value during the recession; more precisely, the depth $D$ of recession $R$ is $D = | min_i ( ADS_i ) |$, $i \in R$, where $i$ denotes days.  Recession severity $S$ is the product of depth and duration. Both $D$ and $S$ use a late-vintage ADS chronology and the NBER recession chronology.
	\end{spacing}
	\end{table}

\subsection{Performance Assessment} \label{aa3}

Truly credible nowcasting performance assessment requires using \textit{vintage information}, which emerges as the limit of a sequence of progressively more realistic and credible nowcast/forecast evaluation approaches:\footnote{Note that nowcasts are effectively just $h$-step-ahead forecasts with horizon $h {=} 0$.}

\begin{enumerate}[label={(\arabic*)}]
	
	\item Use full-sample estimation, and use final revised data  \label{full}
	
	\item Use expanding-sample estimation, and use final revised data   \label{exp}
	
	\item Use expanding-sample estimation, and use vintage data  (``Pseudo Real Time")  \label{vint}
	
	\item Use expanding-sample estimation, and use vintage information  (``Real Time").   \label{info}
	
\end{enumerate}	
Approaches \ref{full} and \ref{exp} are clearly unsatisfactory:  Approach \ref{full}  uses time periods and data values  not available in real time, and approach \ref{exp}  is an improvement but still uses data  values not available in real time. Approach \ref{vint}, involving vintage \textit{data},  is typically viewed as the gold standard.  It is implemented comparatively infrequently, however, due to the tedium involved and the fact that vintage data are often unavailable.\footnote{The two key sources of U.S. vintage data are the Real-Time Dataset for Macroeconomists at the Federal Reserve Bank of Philadelphia (\url{https://www.philadelphiafed.org/research-and-data/real-time-center/real-time-data/}), and ALFRED at the Federal Reserve Bank of St. Louis (\url{https://alfred.stlouisfed.org/}).}  Approach \ref{info},  involving vintage \textit{information}, limits the information  set to that available and actually used  in real time, which is more restrictive than merely limiting the \textit{data} to that available in real time.  It is, however, almost never implemented.

To appreciate why  fully-credible assessment requires vintage information rather than just vintage data, consider the following:  
\begin{enumerate}[label={(\arabic*)}]
	
	\item Econometric/statistical theory and experience  evolve, prompting changes to the estimation procedure; the frequency and timing of  re-estimation and its interaction with benchmark revisions; the estimation sample period; allowance for parameter variation and breaks; the treatment of outliers; the strength of regularization employed; the predictive loss function employed; etc.
	
	\item Economic theory and empirical economic experience  evolve.  Over time this may prompt, for example, the removal or re-weighting of some component nowcast indicators and/or addition of others (e.g., \cite{DR1991}), as well as deeper changes in the nowcasting model.

	\item Exact times and reliability of nowcast/forecast calculation and release may differ due to technological problems; outright mistakes in nowcast/forecast construction; evolving or changing software algorithms and associated bugs; parallel problems at the agencies responsible for the underlying data and decisions regarding how to deal with them in forecast/nowcast construction; etc.

\end{enumerate}
For these and other reasons, just as truly credible evaluation requires refraining from endowing agents with better data than were actually  available in real time, so too does it require refraining from endowing them with better economic or statistical models and related tools than were actually  available in real time, better judgment and decision-making abilities/choices than were  actually manifest in real time, etc.

The upshot is clear:  Truly credible real-time evaluation -- that is, evaluation using vintage information rather than just vintage data -- can only be obtained by using nowcasts produced and permanently recorded in real time.   ADS has been produced and recorded in real time since late 2008, so   I  can credibly study the key episode of current interest,  the Pandemic Recession.   I  now proceed to do so.

\section{The Pandemic Recession Entry and Exit}   \label{PR}

\begin{figure}[tbp]
	\caption{ADS Index:  Ex Post Path  12/1/2007 - 6/26/2020 (Vintage 6/26/2020)}
	\begin{center}
		\includegraphics[trim={40mm 20mm 0mm 25mm},clip,scale=.29]{{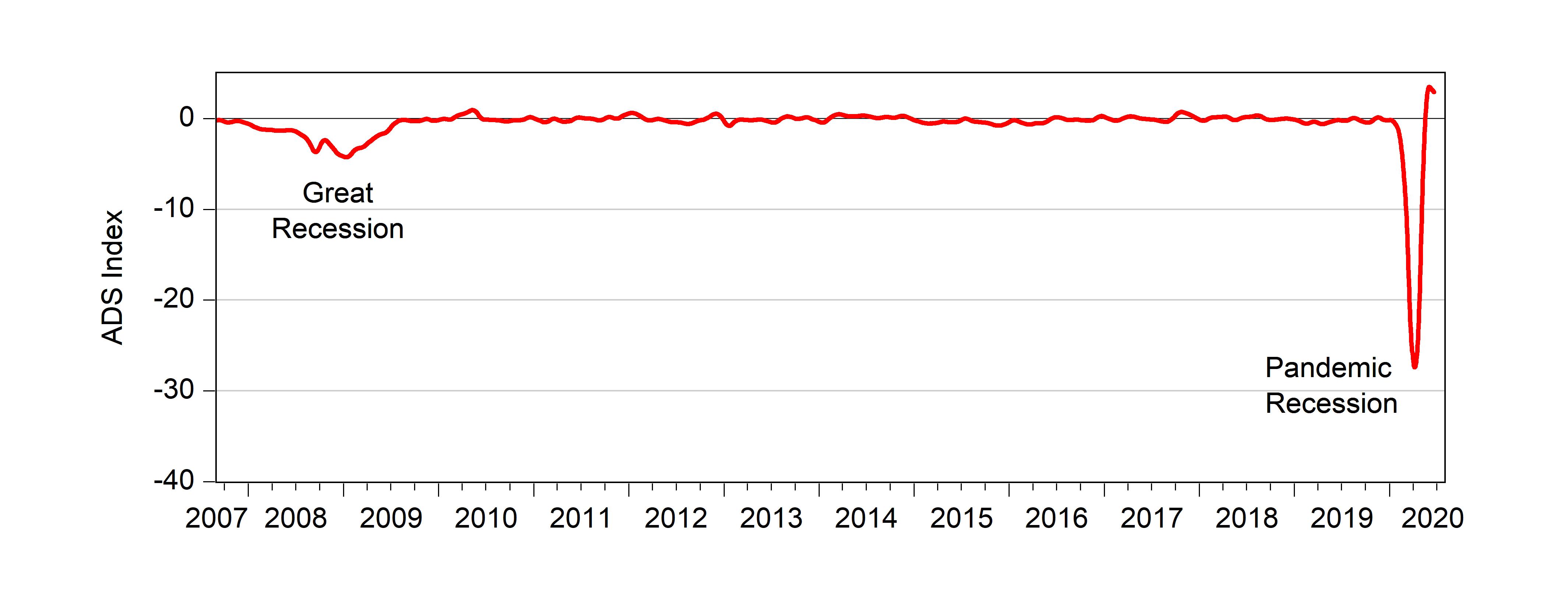}}
	\end{center}
	\label{fig2bbbxxx}
	\end{figure}

  I  focus in this section on the Pandemic Recession of 2020.  It is instructive to begin by comparing it to the Great Recession of 2007-2009.  To that end   I  show the ADS path  in Figure \ref{fig2bbbxxx}, from late 2007 through June 2020.\footnote{ I  refer to an ADS extraction as a  path.}  The so-called ``Great Recession" appears minor  by comparison.

\subsection{A Detailed Look at the Later-Vintage Path}

Figure \ref{fig2bbbxxx} reveals the jaw-dropping ADS drop in the Pandemic Recession, more than five times that of any other recession since 1960.  The ADS drop is entirely appropriate,  due to similarly jaw-dropping and historically unprecedented movements in its underlying indicators.\footnote{See Appendix \ref{keydates} for an annotated chronology of data releases and associated ADS movements.}

\begin{figure}[tbp]
	\caption{ADS Index:  Ex Post Path  1/1/2020 - 6/26/2020 (Vintage 6/26/2020)}
	\begin{center}
		\includegraphics[trim={70mm 45mm 0 50mm},clip,scale=.17]{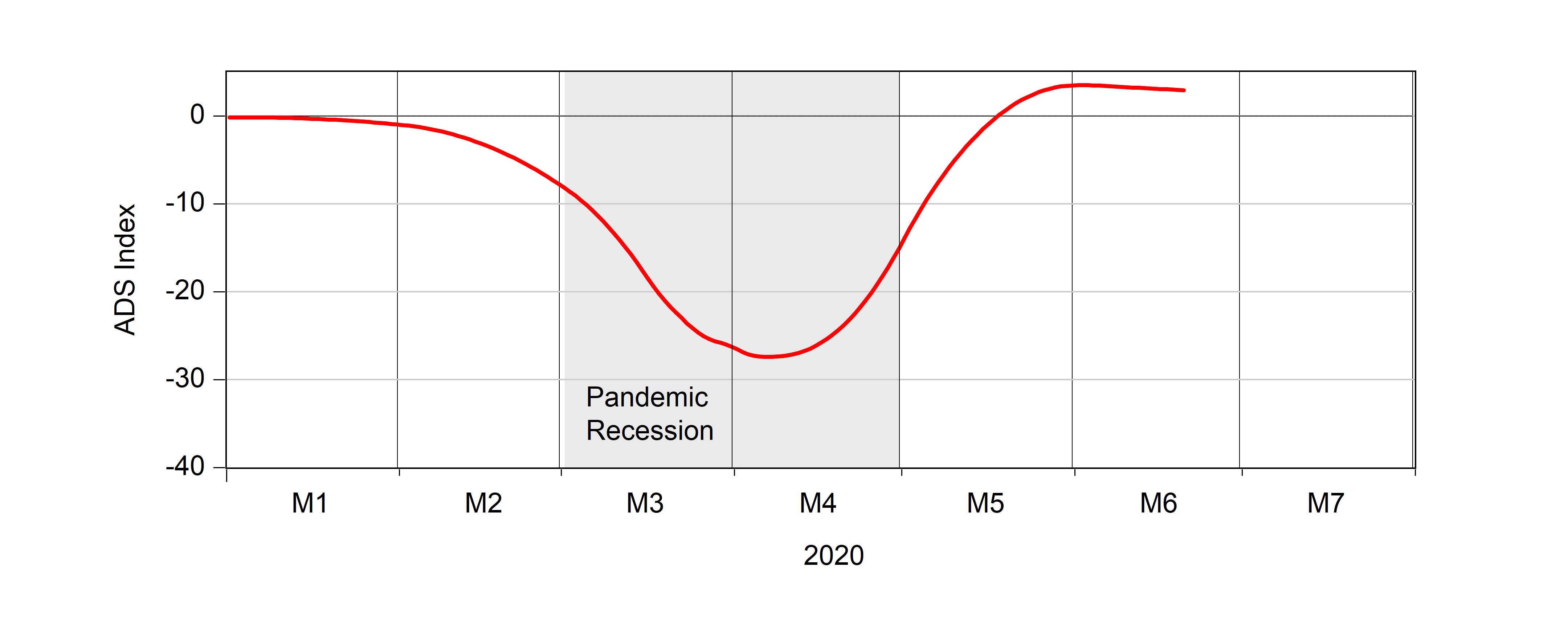}
	\end{center}
	\label{lastfull}
\end{figure}

The official NBER trough month for the Pandemic Recession is April 2020, making the Pandemic Recession the shortest in history.\footnote{Note that although the NBER peak and through months are February and March, respectively,  peaks are allocated to expansions and troughs are allocated to recessions.  Hence the Pandemic Recession duration is two months (March-April 2020), as per Table \ref{NBERreplica}.  Note also that the April 2020 ending date was not determined and announced by the NBER until July 2021.}  In Figure \ref{lastfull},  I  show the later-vintage Pandemic Recession path. The overall extracted path is smooth and convex, with a minimum in early April, and a return to positive growth by early May.\footnote{I  emphasize again, however, that ADS, like the NBER recession chronology, tracks real activity \textit{growth}, not level.  Hence positive ADS does not necessarily mean ``good times"; rather, it means ``good growth", which may be from a very bad initial condition. That was the situation in May, as the battered U.S. economy resumed growth.}  Note therefore that ADS would date the Pandemic Recession's end as May rather than the NBER's April.  I stand by ADS, but the timing difference is of course negligible.  The key difference is that ADS chronologies are available much earlier than the NBER chronology.  The Pandemic Recession chronology shown in Figure \ref{lastfull}, for example, was available on 6/26/2020, whereas the NBER did not release its Pandemic Recession ending date until 7/19/2021.

\subsection{Real-Time Vintages}

\subsubsection{Five Snapshots}

\begin{figure}[tbp]
	\caption{Entering and Exiting  the Pandemic Recession: Monthly Real-Time  ADS Paths}
	\begin{center}
{\includegraphics[trim={75mm 100mm 0 110},clip,scale=.17]{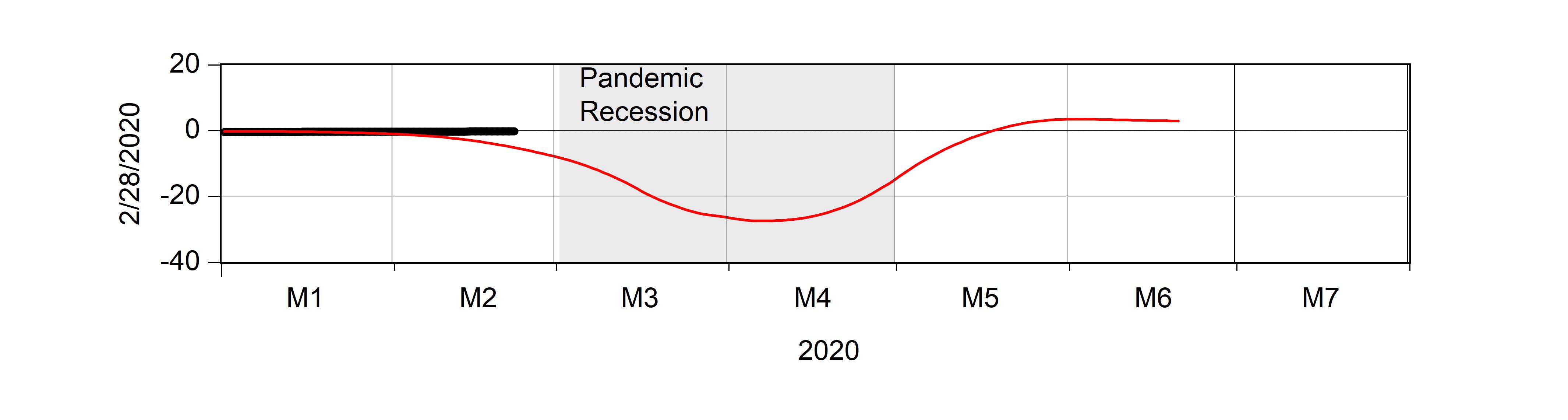}}
{\includegraphics[trim={75mm 100mm 0 50},clip,scale=.17]{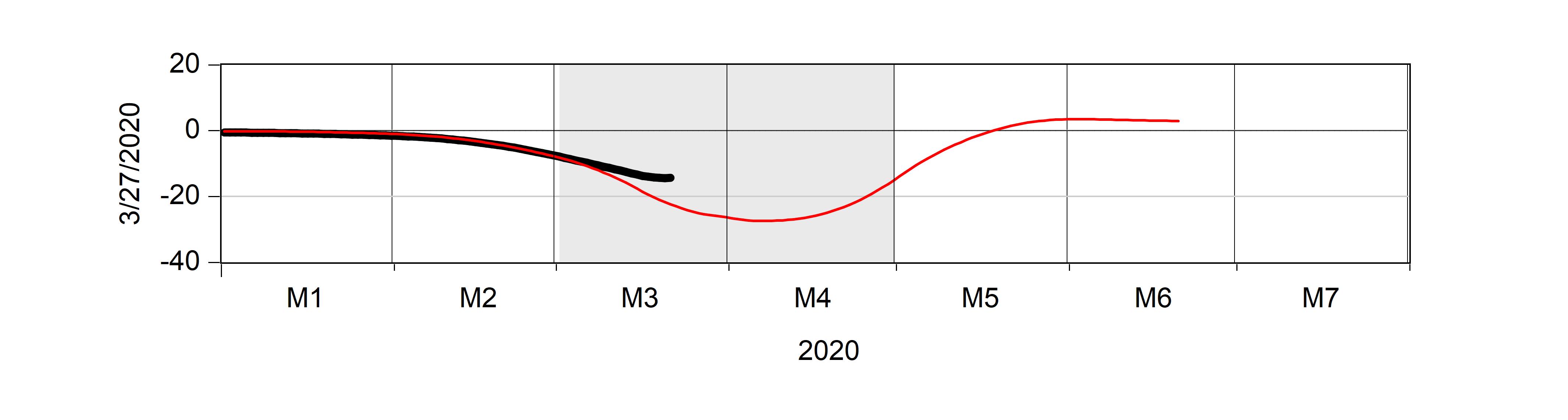}}
{\includegraphics[trim={75mm 100mm 0 50},clip,scale=.17]{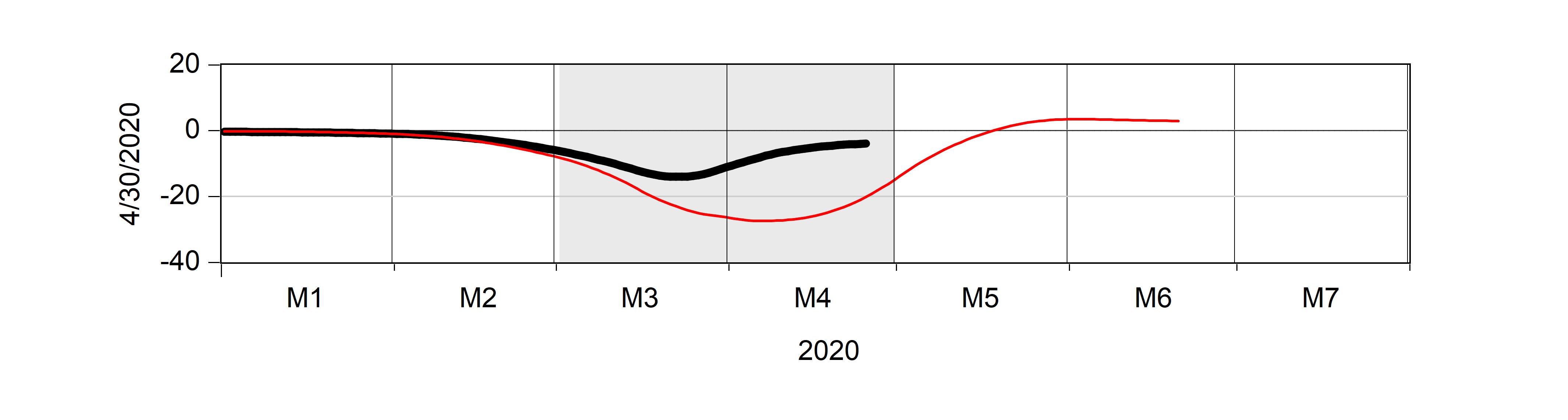}}
{\includegraphics[trim={75mm 100mm 0 50},clip,scale=.17]{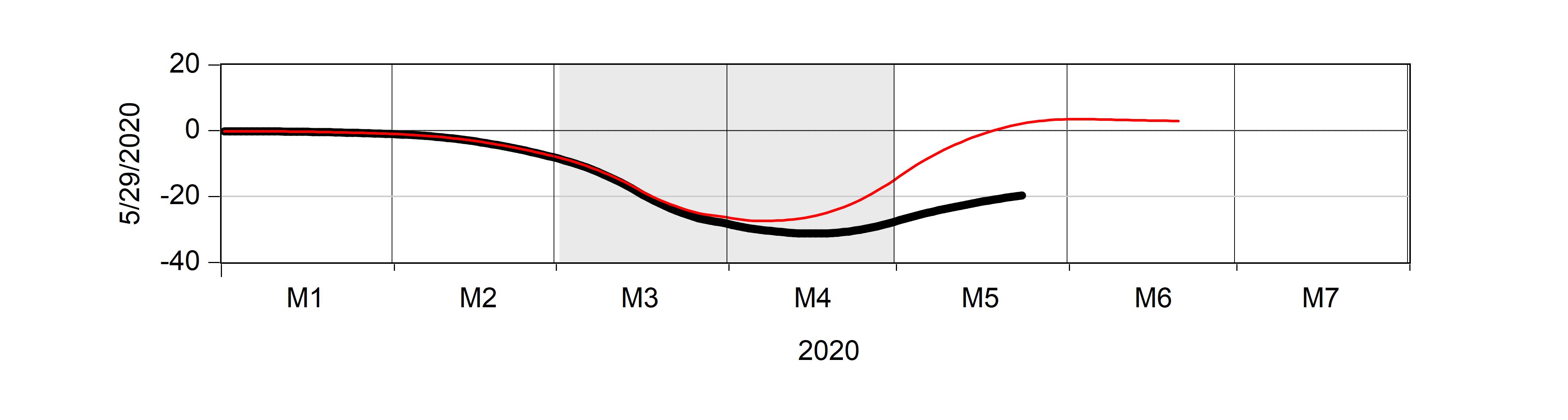}}
\href{https://www.sas.upenn.edu/~fdiebold/papers/Slider_2007_2011.html}{\includegraphics[trim={75mm 30mm 0 50},clip,scale=.17]{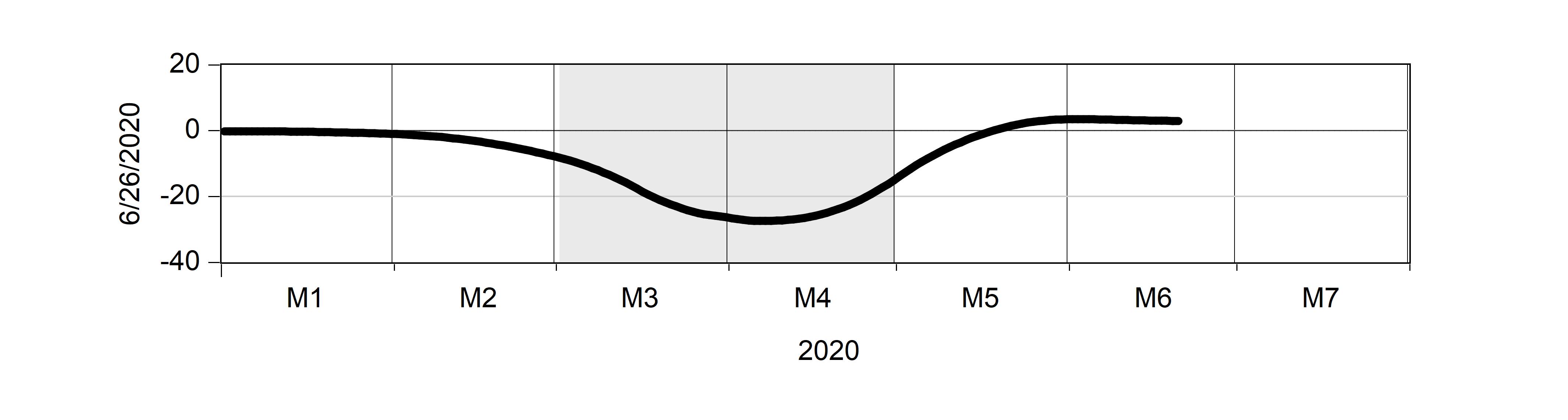}}
	\end{center}
	\label{fig11a}
	Notes:   I  show five monthly real-time ADS paths in black.  From top to bottom they are 2/28/2020, 3/27/2020, 4/30/2020, 5/29/2020, and 6/26/2020.  For comparison  I  show the 6/26/2020 path in red in all panels. (In the bottom panel,  I  show only black, since black and red are identical.) 
\end{figure}

In Figure \ref{fig11a}  I  show several end-of-month paths in black, starting with February 2020.     For comparison, in each panel   I  also show the later-vintage path in red. Moving through the five panels of  Figure \ref{fig11a}:

\begin{enumerate}[label={(\arabic*)}]

\item In the top panel  I  show the 2/28/2020 path.  ADS has not moved.

\item In the second panel  I  show the 3/27/2020 path, which looks very different.  ADS has become acutely aware of the disastrous situation; indeed most of the 3/27 path is well below the previous all-time (post-1960) ADS low during the 1970s oil-shock recession.\footnote{It is also apparent that the Kalman smoother may be smoothing ``too much", producing low ADS values well before mid-March, going back into February and even January.  Its smoothing is optimal relative to the patterns in historical data, but the March initial jobless claims movements were unprecedentedly sharp.} 

\item In the third panel  I  show the 4/30/2020 path.  The April initial claims news is bad, but less bad than March, which is good, and ADS shows a minimum in late March followed by  a rise  toward normalcy by the end of April. 

\item In the fourth panel  I  show the 5/29/2020 path.  The May news is very bad, dominated by the shockingly bad May 8 payroll employment number (for April), and  the late-May path is massively down-shifted relative to the late-April path.  The new minimum is in mid-April rather than late March, and the 5/29 ADS value is thoroughly dismal, nowhere near normalcy.

\item In the fifth panel  I  show the 6/26/2020 path.  Thanks to the strong May payroll employment number (released June 5), ADS moved into normal territory, and stayed there.  There is clear evidence for a Pandemic Recession trough in early/mid May, when ADS hits 0.

\end{enumerate}

\subsubsection{The Full Path Plot and Dot Plot}

\begin{figure}[t]
	\caption{Entering and Exiting the Pandemic Recession: Real Time ADS Path Plot}
	\begin{center}
		\includegraphics[trim={70mm 55mm 0 40mm},clip,scale=.16]{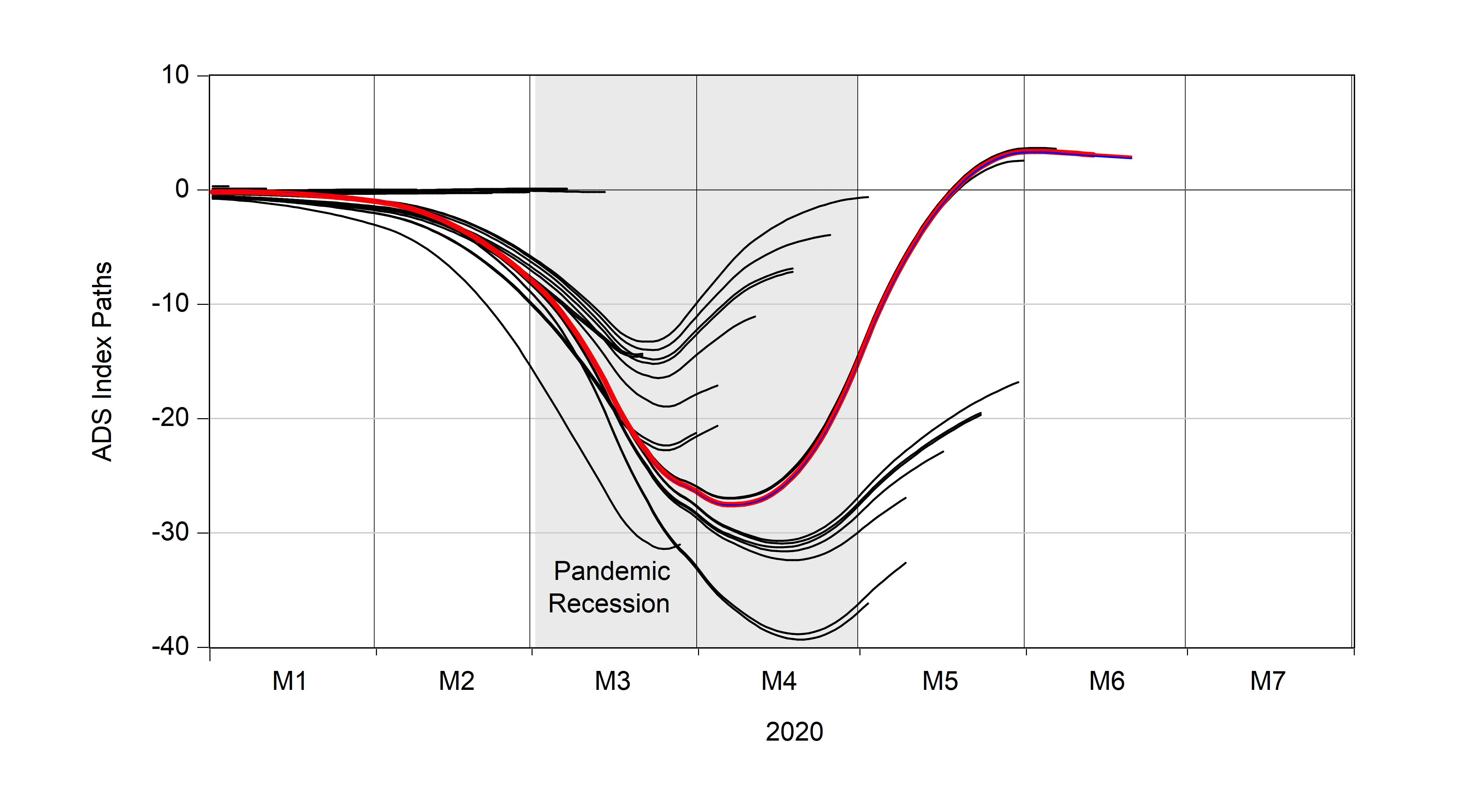}
	\end{center}
	\label{fig2c} 
	Notes:  I  show all real-time  ADS paths in black, through 6/26/2020.  For comparison  I  show the complete later-vintage path (6/26/2020)  in red.  
\end{figure}

In Figure \ref{fig2c},  I  show the complete path plot   during the Pandemic Recession through  6/26/2020, with the  later-vintage path in red for comparison.  The path plot is the set of all real-time paths; by following rightward through the sequence of paths, moving through time,  I  track the evolution of ADS beliefs about the chronology of business conditions.  In Appendix \ref{keydates}  I  provide a corresponding  annotated path chronology. 

There are wide real-time divergences between individual early paths and the later vintage red path.  There are interesting patterns, however, with several real-time ``meta paths" evident:

\begin{enumerate}[label={(\arabic*)}]

\item  The first extends through the 3/19/2020 ADS announcement.  ADS does not move.  Initial claims rise from 0.2m to 0.3m, a large move by historical standards, confirming what everyone already knew: the pandemic would have important real economic consequences, but the Kalman smoother optimally but erroneously ascribes it to measurement error.

\item The second meta-path begins with the 3/26/2020 and 4/2/2020 initial claims explosions.  ADS plunges, but then recovers steadily despite a steady stream of bad news (it is bad, but getting less bad), almost back to 0 by the 5/7/2020 initial claims announcement.  

\item The third meta-path begins with the horrific  5/8/2020 April payroll employment release, with ADS again plunging.  It then again begins mean reverting, and does so completely when the strong May payroll employment number is released on 6/5/2020.   

\end{enumerate}

\begin{figure}[t]
	\caption{ Entering and Exiting  the Pandemic Recession: Real-Time ADS Dot Plot}
	\begin{center}
		\includegraphics[trim={70mm 50mm 0 45mm},clip,scale=.16]{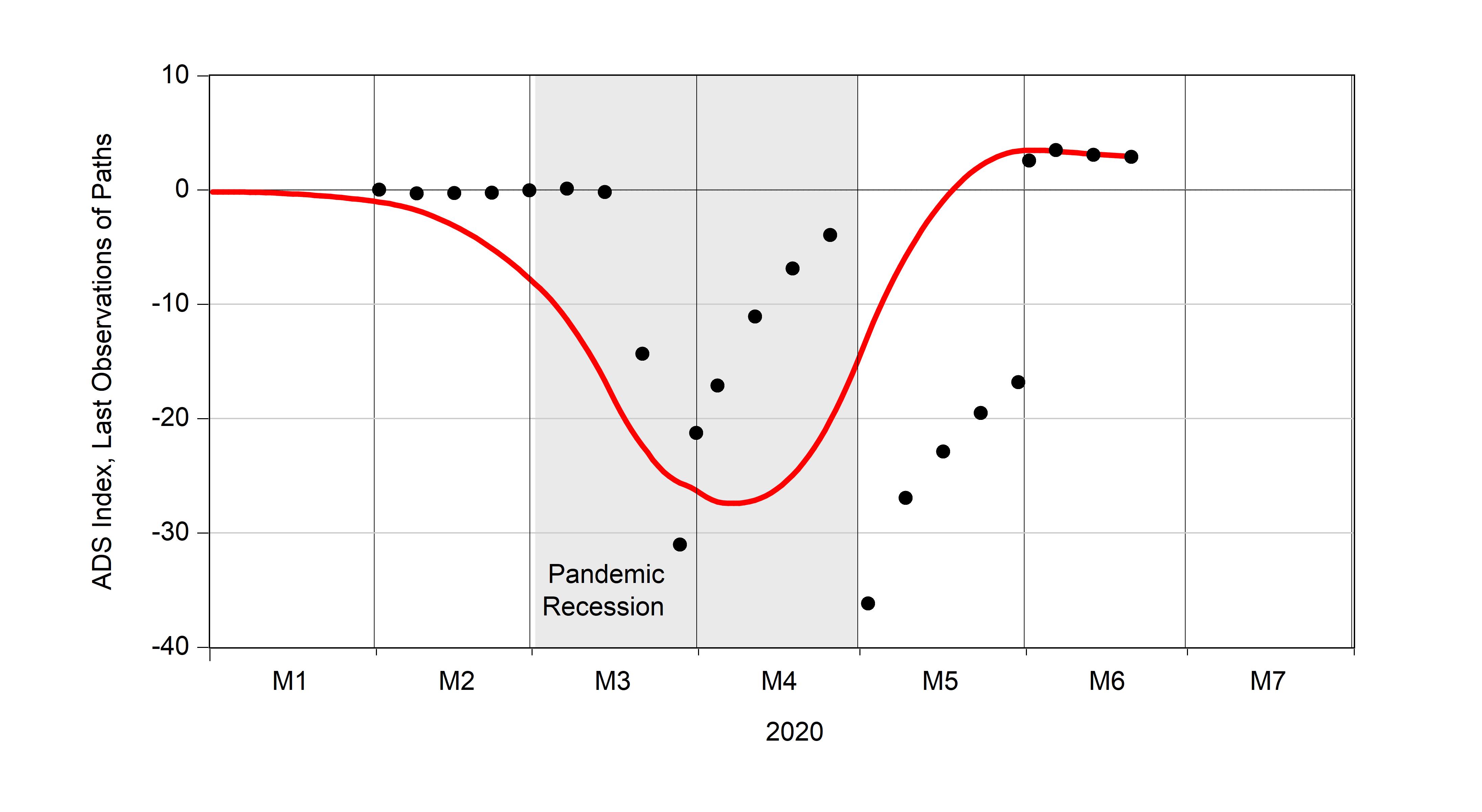}
	\end{center}
	\label{fig2bbc}
	Notes:  I  show the last values of all real-time  ADS paths in black.   For comparison  I  show the complete later-vintage path (6/26/2020)  in red.
\end{figure}

\noindent In Figure \ref{fig2bbc}  I  show the corresponding ``dot plot", with the 6/26/2020 path again superimposed.  Each dot is the last observation of its corresponding path in Figure \ref{fig2c}.  The dots are real-time filtered values, because smoothed and filtered values coincide for the last observation in a sample.   The dot plot is highly volatile and emphasizes the various meta-paths. 

 It is interesting to compare the third ADS (May) real-time meta-path to the late-vintage ADS path, and to the eventually-released NBER chronology, all of which appear in Figure \ref{fig2bbc}. Real-time ADS was gloomy throughout May, even as later-vintage ADS indicates a recovery beginning in early/mid May, and the NBER chronology similarly indicates a recovery beginning by the end of April.  In real time the return to growth was not obvious until the June 5 release of May payroll employment.

\subsection{Real Economic Activity and COVID-19}

\begin{figure}[tbp]
	\caption{Daily ADS and Smoothed Daily COVID-19 Deaths}
	\begin{center}
		\includegraphics[trim={25mm 25mm 0 26mm},clip,scale=.25]{{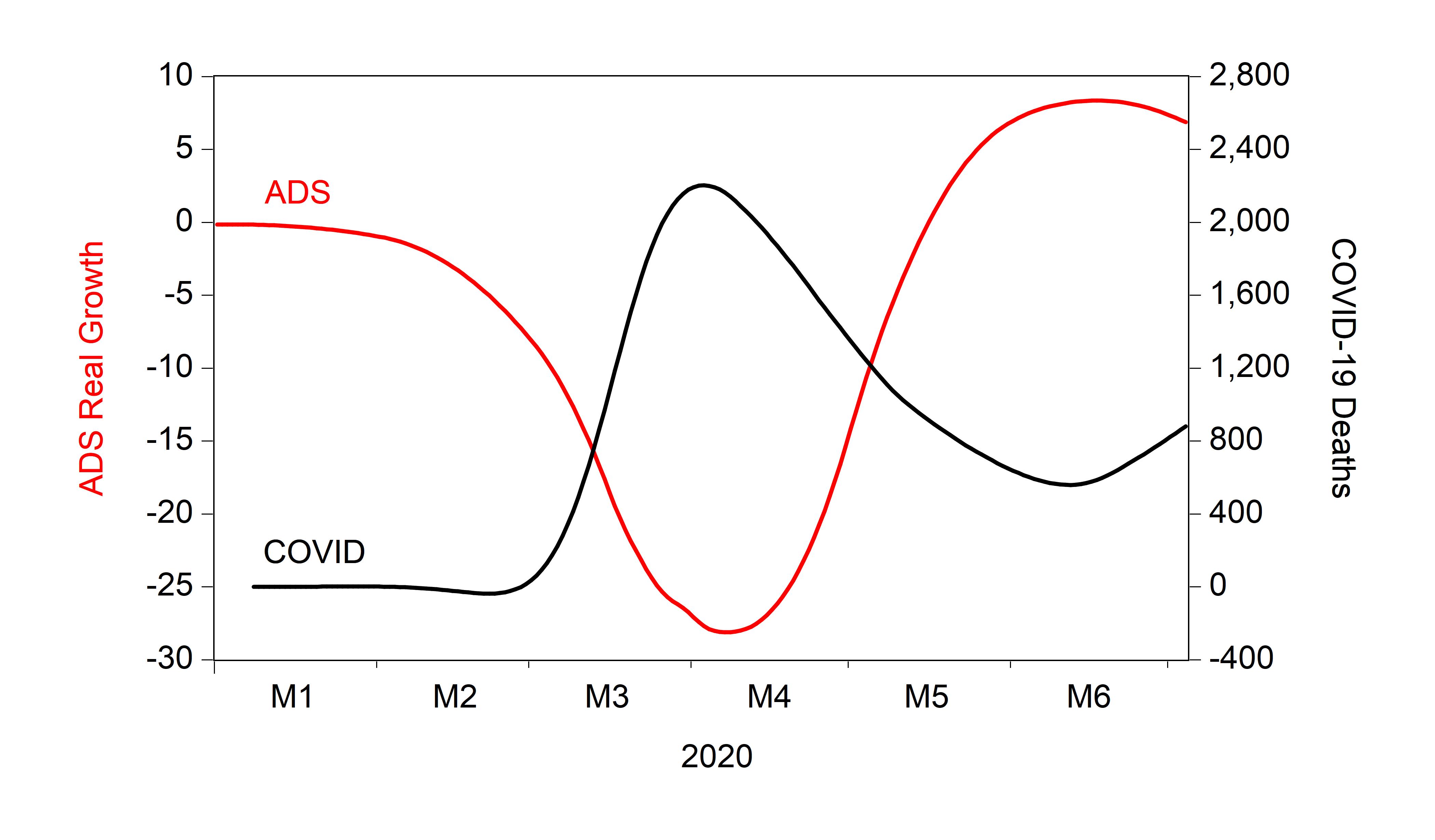}}
	\end{center}
	\label{fig2zz3}
	Notes:  I  show ADS (6/26/2020 vintage) vs HP-filtered daily COVID-19 deaths led by 20 days. See text for details. 
\end{figure}

Because the March-April 2020 collapse in economic activity was obviously caused by COVID-19, it is of interest to directly examine the correlation between the two.   I  can do so at high frequency (daily), because  I  have both daily ADS and COVID new cases / deaths data.   I  want to correlate COVID new cases with ADS, but the direct new cases data are less reliable than deaths during the period of interest, because new cases were likely heavily influenced by changes in the amount of testing undertaken.  Instead, a more reliable indicator of new cases is deaths, adjusted for the approximate 20-day period between infection and death. Hence  I  use deaths led by 20 days.\footnote{ I  use the Johns Hopkins University CSSE COVID-19 daily deaths data; see \cite{Dongetal} and \url{https://github.com/CSSEGISandData/COVID-19}.} In Figure \ref{fig2zz3}  I  show ADS vs COVID deaths+20.\footnote{ I  also smooth COVID deaths+20 using a Hodrick-Prescott filter to remove the strong calendar effects in recorded deaths.} The strength of the negative correlation is striking.  Of course economic activity plunged in March when COVID exploded, but there's much more than that -- ADS and COVID continue to move in lockstep (inversely) through the April COVID peak, its April-May decline, and its June rebound.

\section{Comparison to the Great Recession Exit}  \label{GR}

\begin{figure}[tp]  
	\caption{Exiting the Great Recession: Five Quarterly Real-Time  ADS Paths}
	\begin{center}
		{\includegraphics[trim={35mm 60mm 0 65},clip,scale=.29]{{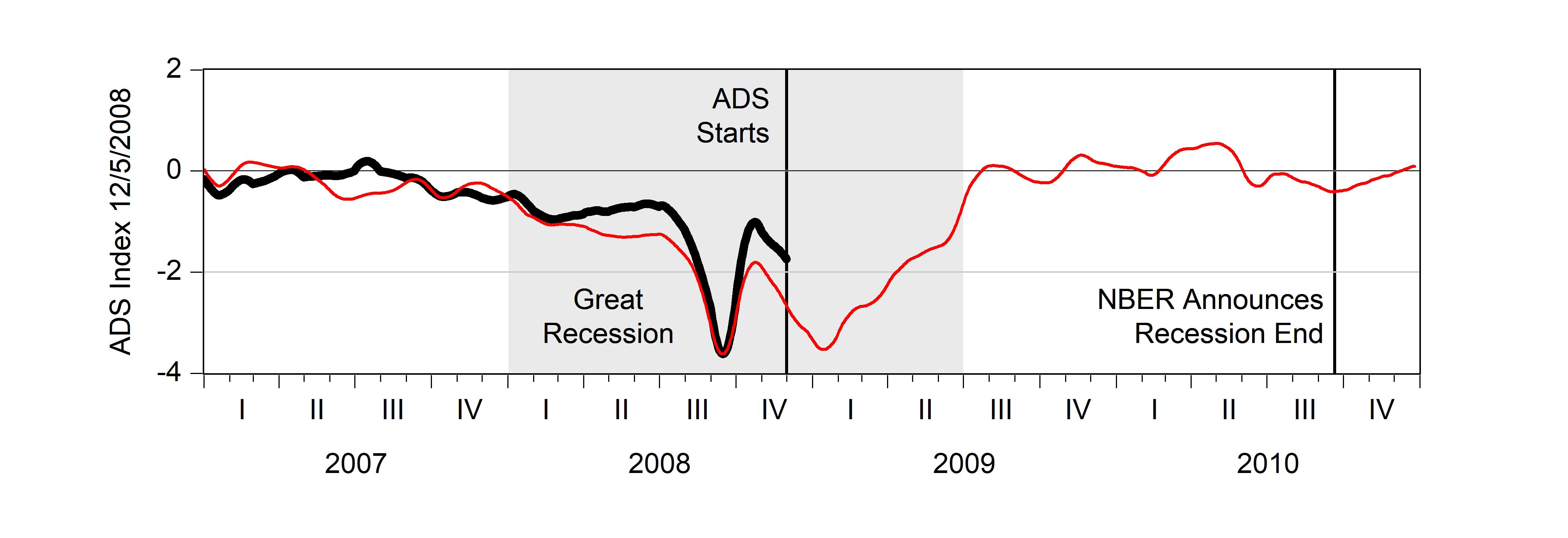}}}
		{\includegraphics[trim={35mm 60mm 0 65},clip,scale=.29]{{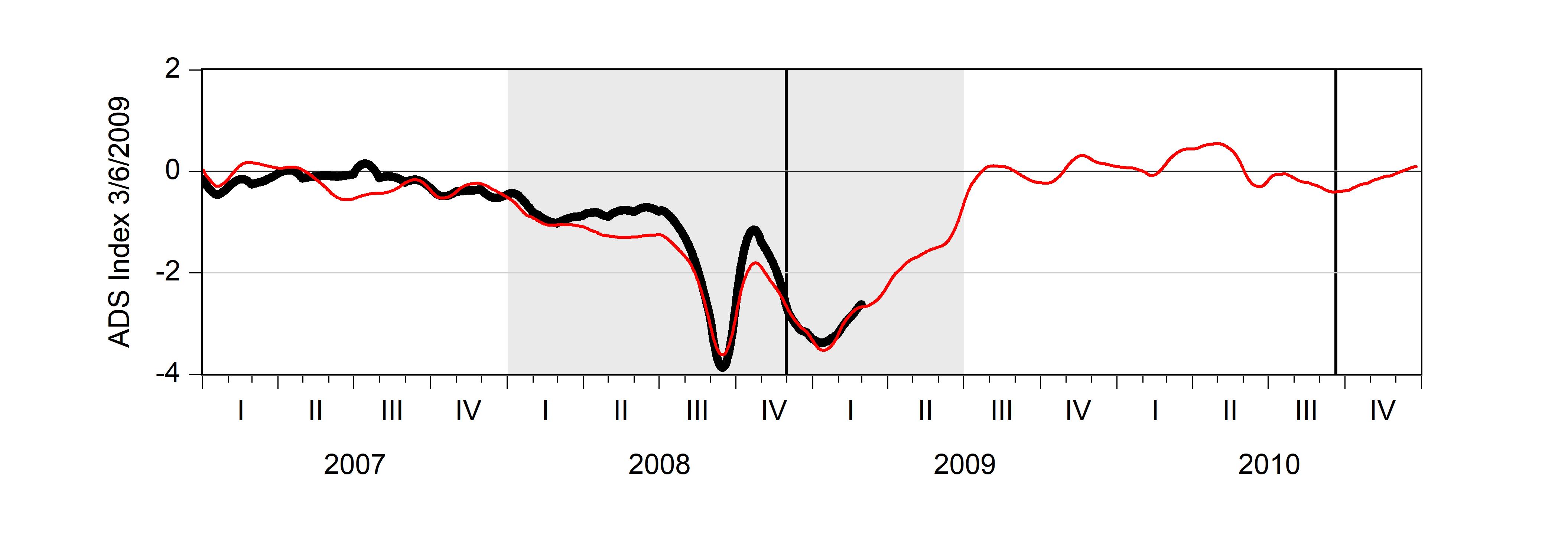}}}
		{\includegraphics[trim={35mm 60mm 0 65},clip,scale=.29]{{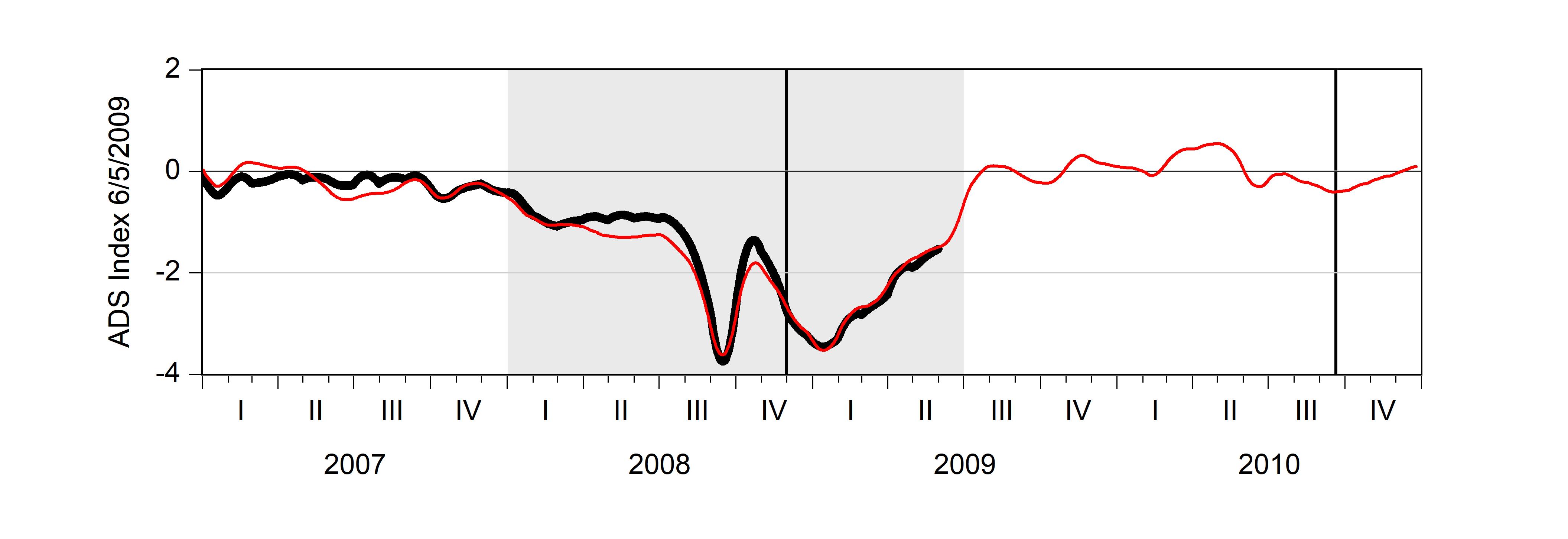}}}
		{\includegraphics[trim={35mm 60mm 0 65},clip,scale=.29]{{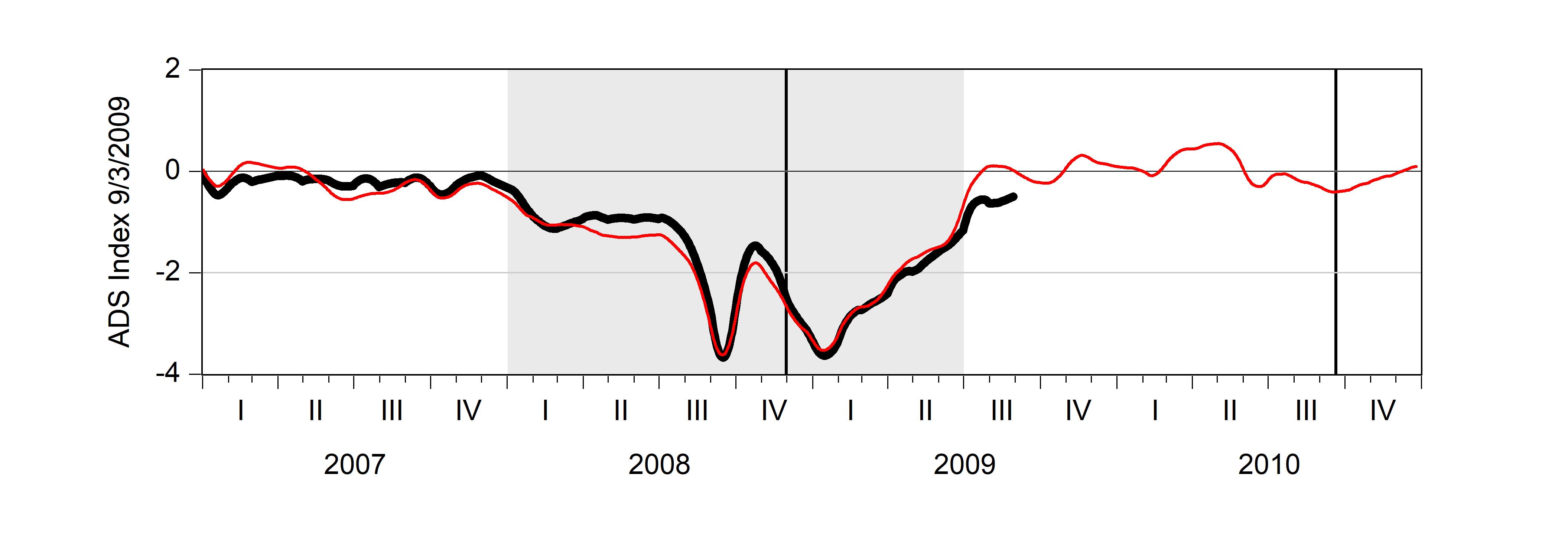}}}
		{\includegraphics[trim={35mm 25mm 0 65},clip,scale=.29]{{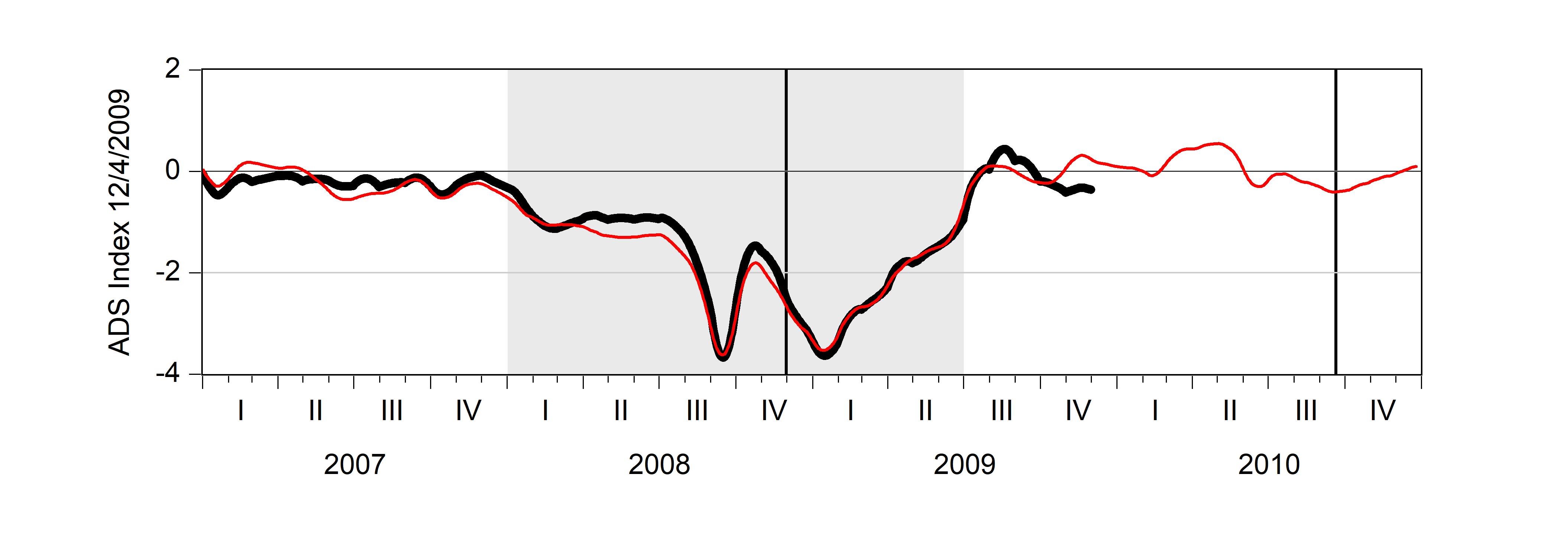}}}
	\end{center}
	\label{fig1a}
	Notes:   I  show five quarterly real-time ADS paths in black.  From top to bottom they are 12/5/2008, 3/6/2009, 6/5/2009, 9/3/2009, 12/4/2009.  For comparison I show a later-vintage ADS path (December 2010)  in red. 
\end{figure}

It is informative to compare the evolution and congealing of views during the Pandemic Recession to those during an earlier, more ``standard", recession, like the Great Recession of 2007-2009.    I  can't examine real-time ADS  when entering the Great Recession, because ADS did not start until December 2008, well after the great recession began. But  I  can examine it  when \textit{exiting} the great recession. In Figure \ref{fig1a}  I  show five paths in black, from ADS inception through the end of the Great Recession, at quarterly intervals.  For comparison  I  also show a later-vintage path in red. It is revealing to  compare the real-time paths to the ex post path.   One can think of the later-vintage path as ``truth", or at least a good  assessment of truth based on later-vintage data. 
%\footnote{See the interactive graphic (slider and animation) at \url{https://www.sas.upenn.edu/~fdiebold/papers/Slider_2007_2011.html}.} 

In the top panel of Figure \ref{fig1a}  I  show the first ADS path, 12/5/2008.  ADS shows a very deep recession, almost the deepest on record since 1960, bottoming out in 2008Q3, with movement toward recovery in late Q3 and early Q4, even if it had stalled a bit by early December.  As it turned out, however, the  Great Recession subsequently featured a growth rate ``double dip".  The 12/5/2008 ADS path ends just after the first dip, which involved a sharp drop in September 2008 and an equally sharp rebound.\footnote{In particular, according to the Federal Reserve’s G.17 Industrial Production (IP) release of October 16, 2008, September IP was severely affected by a highly‐unusual and largely exogenous “triple shock” (Hurricanes Gustav and Ike, and a strike at a major aircraft manufacturer), which caused an
	annualized September IP drop of nearly fifty percent.  A similar pattern exists for Manufacturing and
	Trade Sales (MTS).  IP and MTS also rebounded unusually sharply in October – indeed IP appears to “overshoot” –
	presumably in an attempt by manufacturers to make up for September’s loss.
}  At the time it was easy to read the cards as saying that the recession was ending, and  ADS was a bit too optimistic, moving upward toward recovery. 

Now consider the remaining  panels of Figure \ref{fig1a}.  In the second panel   I  show the next, and contrasting, 3/6/2009 ADS path.  In the interim ADS has quickly learned the situation, the double dip in particular, and is very much on track,  capturing the second dip in January 2009. ADS continues to climb steadily through the third and fourth panels (6/5/2009 and 9/3/3009, respectively), and by the time of the bottom panel (12/4/2009) it is clear that the Great Recession ended in June or July, with ADS basically fluctuating around 0 after that.  (Recall that ADS${=}$0 means average growth, not zero growth.)  All told,  the five quarterly real-time ADS paths generally  match the ex post path closely, and they correctly identify the recession's  end, well before the end of 2009 and indeed roughly 1.5 \textit{years} before the official NBER announcement in September 2010.  

To emphasize ADS timeliness,   I  plot the later-vintage ADS in Figure \ref{fig1a} all the way through 2010, which allows inclusion of the NBER's end-of-recession announcement on 9/20/2010, long after the fact and not helpful for real-time decision making.\footnote{In fairness it must be noted that the NBER is not \textit{seeking} to be helpful for real-time decision making; rather, it seeks to meticulously construct the U.S. business cycle chronology of record, quite reasonably  using all relevant information -- even very late-arriving information.}   ADS fills the gap left by the late-arriving NBER chronology, and it also provides a numerical measure that allows one to track the recession's pattern, depth, overall severity, etc., in addition to duration.\footnote{I  could have included an end-of-recession marker in Figure \ref{fig11a} for the Pandemic Recession as well, but doing so would have distorted the figure beyond recognition, given that the recession lasted only two months and the announcement came well over a year later.}

\begin{figure}[tbp]
	\caption{Exiting the Great Recession:  Real-Time ADS Path Plot}
	\begin{center}
		\includegraphics[trim={25mm 25mm 0 26mm},clip,scale=.25]{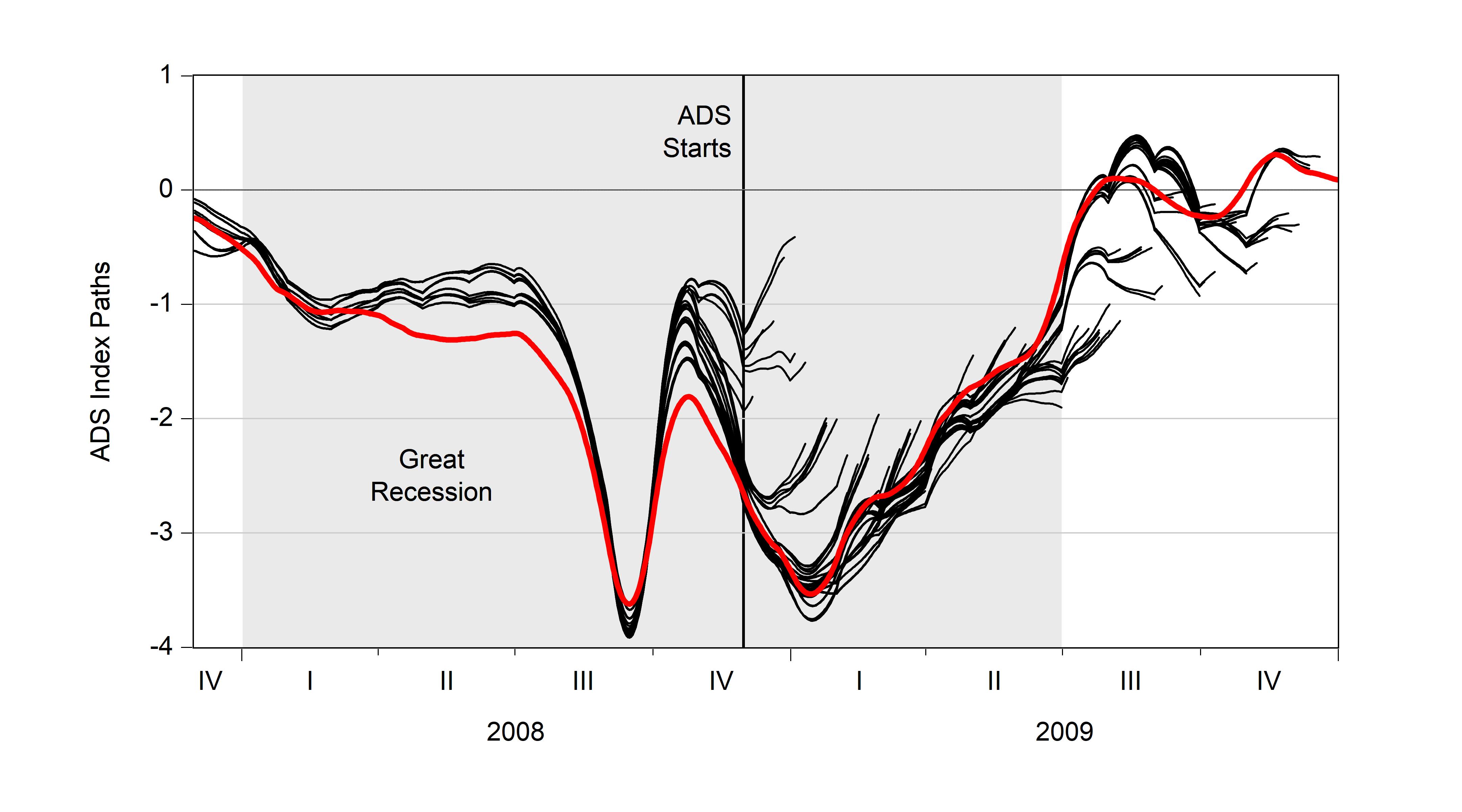}
	\end{center}
	\label{fig2aa}
	Notes:  I  show all 2008-2009 ADS paths since the first on 12/5/2008.    I  show real-time ADS paths in black, and a comparison late-vintage ADS path (December 2010) in red.   
\end{figure}

\begin{figure}[tbp]
	\caption{ Exiting the Great Recession: Real-Time ADS Dot Plot}
	\begin{center}
		\includegraphics[trim={25mm 25mm 0 26mm},clip,scale=.25]{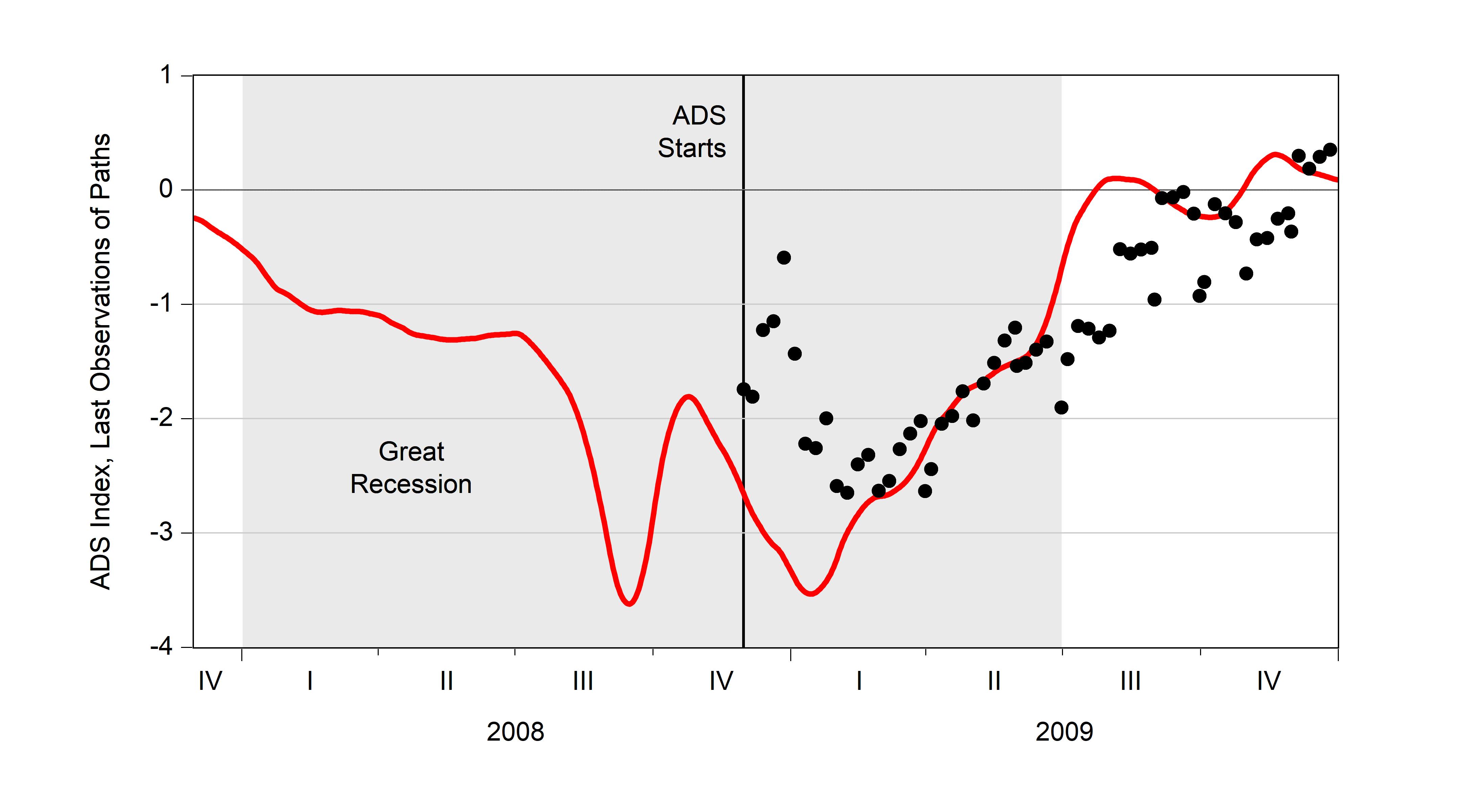}
	\end{center}
	\label{fig2bb}
	Notes:  I  show the last values of each 2008-2009 ADS path in black, with a comparison later-vintage ADS path (December 2010) superimposed in red.
\end{figure}

For example, and as recorded in Table \ref{NBERreplica}, ADS identifies the Great Recession as the worst since 1960, with longest duration and third-greatest depth, resulting in the greatest overall severity (duration times depth).  The Pandemic Recession, in contrast, was the deepest on record by an order of magnitude, but it was also  the shortest by far, attaining a rank of  third in overall severity (behind the 2007-2009 Great Recession and the 1973-1975 Oil Shock Recession). 

In Figure \ref{fig2aa}  I  show the complete path plot.   Of course there are errors positive and negative as the recession evolves, but overall ADS performs well, sending a reliable and valuable signal for navigating the path out of recession.  I  show the corresponding dot plot in Figure \ref{fig2bb}.  The dots are real-time smoothed values.\footnote{They are also filtered, because smoothed and filtered values coincide for the last observation in a sample.}  The black-dot sequence of real-time smooths is  naturally less variable than the later-vintage (December 2010) smooth shown in red, because the latter has  more information on which to condition, and therefore  captures more variation.

\section{Concluding Remarks and Directions for Future Research}  \label{concl}

Our approach was part methodological and part substantive.  On the substantive side,  I   explored how views formed using the ADS  nowcast  evolved when entering/exiting the U.S. Pandemic Recession, which arrived abruptly and then ended quickly.  In particular,  I  tracked the evolution of real-time vintage beliefs and  compared them to a   later-vintage  chronology.  ADS real activity growth plunged wildly in March 2020 and swung in real time as its underlying components swung, but it returned to brisk growth by mid May.    I  also documented a strong negative relationship between the real-time ADS Pandemic Recession  path and the concurrent real-time COVID-19  path, and  I  compared aspects of the  Pandemic Recession and Great Recession paths.

On the methodological side,  I  clarified the meaning of truly honest real-time nowcast/forecast evaluation and illustrated it using the ADS Business Conditions Index, which has now been in operation over a long span that includes  emergence from the Great Recession, entry into the Pandemic Recession, and exit from the Pandemic Recession.  An interesting methodological direction for future work would be decomposition of ADS movements into shares coming from the underlying indicators.  One approach would be to use the observational weights implicit in the Kalman filter, as in  \cite{koopman2003computing}.  Another approach, popular in the recent machine learning literature, would be based on \cite{shapley1953value} values, as in \cite{israeli2007shapley}.  For initial explorations and insights, see \cite{Liu2021}.

%An obvious remaining question of interest is how the Pandemic Recession will look -- when all the smoke clears -- in comparison to its ancestors.  In my view as of April 2021, the smoke has already cleared, and the recession ended long ago.  The NBER  chronology does not measure recession deepness, but ADS does, and it reveals that the Pandemic Recession is clearly the deepest U.S. recession since 1960.  The NBER chronology \textit{does} measure recession duration, but as of April 2021 the NBER  has not announced the Pandemic Recession's ending date, so  I  don't know its ``official" duration.  Nevertheless  ADS clearly indicates a return to sustained positive growth by mid-May 2020, which would make the Pandemic Recession not only  the deepest recession at least since 1960, but also the all-time shortest, by a wide margin (presently the shortest is the six-month recession of early 1980).  Any eventual claim that it ended later than May 2020 would be more an indication of reluctance to declare such a short recession than an unbiased assessment of economic reality. 
%

\clearpage

\appendix
\appendixpage
\addappheadtotoc
\newcounter{saveeqn}
\setcounter{saveeqn}{\value{section}}
\renewcommand{\theequation}{\mbox{\Alph{saveeqn}.\arabic{equation}}} \setcounter{saveeqn}{1}
\setcounter{equation}{0}

\section{Annotated ADS Chronology, 3/17/2020-7/2/2020}  \label{keydates}

[Selected annotations, associated with large real-time ADS moves, appear in italics.]

\bigskip

\noindent 03/17/2020, Industrial Production (for February 2020), data 09:15, ADS 10:45  

\smallskip 

\noindent This is the day of the last ADS update  before the March 19 initial claims release.  ADS continued its more-or-less random vibration around zero, sending the same signal that it had sent since the end of the Great Recession in 2009: the economy is growing normally.   ADS${=}$0.1.

\bigskip

\noindent 03/19/2020, Initial Jobless Claims (for week ending 03/14/2020), data 08:30,  ADS 10:00

\smallskip 

\noindent  IJC took a large move upward, suggesting that the pandemic would have important real economic consequences.  The Kalman smoother optimally but erroneously ascribed this  first-time IJC jump almost entirely to measurement error, and ADS basically did not move.  ADS${=}$-0.2.

\bigskip

\noindent 03/26/2020, Initial Jobless Claims (for week ending 03/21/2020), data 08:30, ADS 10:00  

\smallskip

\noindent  \textit{IJC spiked in jaw-dropping off-the-chart fashion.  Two huge  IJC  moves in a row are \textit{not} optimally ascribed  to measurement error by the Kalman smoother; rather, they are naturally ascribed to the underlying serially-correlated real activity factor -- and ADS drops to approximately -15 in similarly (and literally) off-the-chart fashion.  By way of comparison,  the all-time ADS lows since 1960 were in the recessions of 1973-1975 and 2007-2009, in both cases between -4 and -5.  Note that the ADS path now begins its drop earlier in the year, a result of the serial correlation in   IJC  interacting with the Kalman smoother. It is interesting to speculate as to whether real activity really \textit{was} lower in February (say), due for example to the virus-induced January-February collapse of a major trading partner (China). ADS${=}$-14.5.} \color{black}

\bigskip

\noindent 03/26/2020, Real GDP (fourth quarter 2019, third release), data 08:30, ADS 10:00  

\smallskip

\noindent     Irrelevant.  ADS=-14.5.   

\bigskip

\noindent 03/27/2020,  Real Manufacturing  \& Trade Sales  (for January 2020), data 08:30, ADS 10:00  

\smallskip

\noindent Irrelevant.  ADS=-14.3.

\bigskip

\noindent 03/27/2020,  Real Personal Income Less Transfers (for February 2020), data 08:30, ADS 10:00  

\smallskip

\noindent Irrelevant.  ADS=-14.3.

\bigskip

\noindent 04/02/2020,  Initial Jobless Claims (for week ending 03/28/2020), data 08:30, ADS 10:00  

\smallskip

\noindent     \textit{IJC doubles off-the-charts, and ADS similarly doubles (downward)  to -31.  The Kalman smoother now has ADS beginning its decline in early January, again presumably an artifact of the serial correlation in   IJC  interacting with the Kalman smoother.  Or, again, perhaps it's real.  ADS${=}$-31.0. \color{black} 
}

\bigskip

\noindent 04/03/2020,  Payroll Employment (for March 2020), data 08:30, ADS 10:00  

\smallskip

\noindent     PE drops but ADS \textit{rises}. ADS evidently views the PE drop as good news, because it's not such a big drop compared to the off-the-charts drops.   ADS${=}$-21.2.

\bigskip

\noindent 04/09/2020,  Initial Jobless Claims (for week ending 04/04/2020), data 08:30, ADS 10:00  

\smallskip

\noindent    Another massive   IJC  increase, but ADS largely unchanged.  ADS${=}$-20.6.

\bigskip

\noindent 04/15/2020, Industrial Production (for March 2020), data 09:15, ADS 10:45  

\smallskip

\noindent    IP plunges, but it's for the previous month, and  ADS actually  continues its gradual upward mean reversion as initial claims improve. ADS${=}$-17.1. 

\bigskip

\noindent 04/16/2020, Initial Jobless Claims (for week ending 04/11/2020), data 08:30, ADS 10:00

\smallskip

\noindent     IJC drops some, and ADS improves.  ADS${=}$-11.1.

\bigskip

\noindent 04/23/2020, Initial Jobless Claims (for week ending 04/18/2020), data 08:30, ADS 10:00  

\smallskip

\noindent       IJC and ADS again improve. ADS${=}$-7.2. 

\bigskip

\noindent 04/29/2020, Real GDP (first quarter 2020, first release), data 08:30, ADS 10:00 

\smallskip 

\noindent      -4.8 \% annualized.  2008Q4 was worse (-7.5 \%) but the 2020Q1 number was driven only by (part of) March.  Had January, February, and early March 2020 been as bad as late March, 2020Q1 GDP growth would have been massively worse, consistent with the massive late-March ADS drop.  ADS is essentially unchanged.  In the absence of the GDP news, ADS would presumably have risen, but the bad GDP news provided an offset. ADS${=}$-6.9.

\bigskip

\noindent 04/30/2020, Initial Jobless Claims (for week ending 04/25/2020), data 08:30, ADS 10:00  

\smallskip

\noindent   IJC continues slowly improving.  ADS${=}$ -3.94.
 
\bigskip
 
\noindent 04/30/2020, Real Manufacturing  \& Trade Sales  (for February 2020) , data 08:30, ADS 10:00  

\smallskip

\noindent  February RMTS is irrelevant.  

\bigskip

\noindent 04/30/2020, Real Personal Income Less Transfers (for March 2020), data 08:30, ADS 10:00  

\smallskip 

\noindent       PILT for the previous month down sharply.  The day's news is all bad, yet not so bad as it was, and ADS improves.  Now it approximately equals  its worst value during the Great Recession. ADS${=}$-3.9.

\bigskip

\noindent 05/07/2020, Initial Jobless Claims (for week ending 05/02/2020), data 08:30, ADS 10:00  

\smallskip 

\noindent     IJC again improving.    The IJC numbers continue to be bad, but they are getting less bad, and ADS seems driven by that. By this time the path plot makes clear that new data are causing sizable revisions in entire paths.  For example, the huge ADS trough was estimated to be approximately -32 in the 4/2 vintage, but it was progressively moved upward in subsequent vintages, and now in the 5/7 vintage it is approximately -12. ADS${=}$-0.6.

\bigskip

\noindent 05/08/2020,  Payroll Employment (for April 2020), data 08:30, ADS 10:00  

\smallskip 

\noindent       \textit{Plunges downward, and ADS plunges similarly to an all-time low.   ADS${=}$-36.2. \color{black} }

\bigskip

\noindent 05/14/2020, Initial Jobless Claims (for week ending 05/09/2020), data 08:30, ADS 10:00  

\smallskip 

\noindent    IJC almost unchanged; ADS improves slightly to ADS${=}$-32.6.

\bigskip

\noindent 05/15/2020, Industrial Production (for April 2020), data 09:15, ADS 10:45  

\smallskip 

\noindent       Plunges but ADS nevertheless ADS improves.  ADS${=}$-26.9.

\bigskip

\noindent 05/21/2020, Initial Jobless Claims (for week ending 05/16/2020), data 08:30, ADS 10:00 

\smallskip 

\noindent   IJC improves, and ADS improves to ADS${=}$-22.9.

\bigskip

\noindent 05/28/2020, Initial Jobless Claims (for week ending 05/23/2020), data 08:30, ADS 10:00 

\smallskip 

\noindent      IJC improves slightly.  ADS${=}$ -19.7.

\bigskip

\noindent 05/28/2020, Real GDP (first quarter 2020, second release), data 08:30, ADS 10:00  

\smallskip 

\noindent    Q1 GDP revised down slightly.   ADS${=}$-19.7.

\bigskip

\noindent 05/29/2020, Real Manufacturing  \& Trade Sales  (for March 2020), data 08:30, ADS 10:00  

\smallskip 

\noindent       Down sharply. 

\bigskip

\noindent 05/29/2020, Real Personal Income Less Transfers (for April 2020), data 08:30, ADS 10:00  

\smallskip 

\noindent       Down sharply.   ADS${=}$-19.5.

\bigskip

\noindent 06/04/2020, Initial Jobless Claims (for week ending 05/30/2020), data 08:30, ADS 10:00  

\smallskip 

\noindent     IJC continues its ever-so-slow reversion to normalcy.  ADS${=}$-16.8.
%
%\bigskip
%
%\textbf{---------- Here begins a positive ADS regime in the 2-3\% range. ---------- }

\bigskip

\noindent 06/05/2020,  Payroll Employment (for May 2020), data 08:30, ADS 10:00  

\smallskip 

\noindent  \textit{PE \textit{increases} sharply.  Finally some good news.  ADS jumps upward and goes positive. ADS${=}$2.6.
	\color{black}
}
\bigskip

\noindent 06/11/2020, Initial Jobless Claims (for week ending 06/06/2020), data 08:30, ADS 10:00 

\smallskip 

\noindent   IJC improves slightly. ADS${=}$3.6.

\bigskip

\noindent 06/16/2020, Industrial Production (for May 2020) , data 09:15, ADS 10:45  

\smallskip 

\noindent        IP up, ADS unchanged.  ADS${=}$3.6.

\bigskip

\noindent 06/18/2020, Initial Jobless Claims (for week ending 06/13/2020), data 08:30, ADS 10:00  

\smallskip 

\noindent   IJC basically unchanged. ADS${=}$3.1.

\bigskip

\noindent 06/25/2020, Initial Jobless Claims (for week ending 06/20/2020), data 08:30, ADS 10:00 

\smallskip 

\noindent IJC basically unchanged. 

\bigskip

\noindent 06/25/2020, Real GDP (first quarter 2020, third release), data 08:30, ADS 10:00 

\smallskip 

\noindent      GDP basically unchanged. ADS${=}$2.8.

\bigskip

\noindent 06/26/2020, Real Manufacturing  \& Trade Sales  (for April 2020),  data 08:30, ADS 10:00  

\smallskip 

\noindent      April MTS was weak as expected.

\bigskip

\noindent 06/26/2020, Real Personal Income Less Transfers (for May 2020), data 08:30, ADS 10:00  

\smallskip 

\noindent     May PILT was strong, returning to positive growth. ADS${=}$2.9.

\bigskip

\noindent 07/02/2020, Initial Jobless Claims (for week ending 06/27/2020), data 08:30, ADS 10:00

\clearpage

\bibliographystyle{Diebold}
\addcontentsline{toc}{section}{References}
\bibliography{Bibliography,citation}

\end{document}